\documentclass[preprint, 12pt, a4paper]{elsarticle}

\usepackage{amsthm}
\usepackage{graphicx} % Required for inserting images
\usepackage{longtable}
\usepackage{booktabs,multirow,tabularx} % for better table rules
\usepackage{multicol} % for multiple columns
\usepackage{float}
\usepackage{amsmath, amssymb, mathtools}
\usepackage{supertabular,booktabs}
\usepackage{makecell}
\usepackage{caption}
\usepackage{setspace}
\usepackage{bm}
\usepackage{url}
\usepackage{hyperref}
\hypersetup{
    breaklinks=true,
}
\usepackage{tikz}
\usepackage{qtree}
\usepackage{tikz-qtree}
\usepackage{subcaption}

\usepackage{multirow}
\usepackage{tcolorbox}
\usepackage{xcolor}
\usepackage{listings}
\usepackage{algorithm}
\usepackage{algpseudocodex}
\algrenewcommand\algorithmicrequire{\textbf{Input:}}
\algrenewcommand\algorithmicensure{\textbf{Output:}}
\renewcommand{\mathbf}[1]{\bm{#1}}
\definecolor{codegreen}{rgb}{0,0.6,0}
\definecolor{codegray}{rgb}{0.5,0.5,0.5}
\definecolor{codepurple}{rgb}{0.58,0,0.82}
\definecolor{backcolour}{rgb}{0.95,0.95,0.92}

\lstdefinestyle{mystyle}{
    backgroundcolor=\color{backcolour},   
    commentstyle=\color{codegreen},
    keywordstyle=\color{magenta},
    numberstyle=\tiny\color{codegray},
    stringstyle=\color{codepurple},
    basicstyle=\ttfamily\footnotesize,
    breakatwhitespace=false,         
    breaklines=true,                 
    captionpos=b,                    
    keepspaces=true,                 
    numbers=left,                    
    numbersep=5pt,                  
    showspaces=false,                
    showstringspaces=false,
    showtabs=false,                  
    tabsize=2
}

\journal{SoftwareX}

\begin{document}

\begin{frontmatter}

\title{TorchCor: High-Performance Cardiac Electrophysiology Simulations with the Finite Element Method on GPUs}

\author[a]{Bei Zhou\corref{cor1}}
\ead{bei.zhou@imperial.ac.uk}
\author[a,b]{Maximilian Balmus}
\author[a]{Cesare Corrado}
\author[a]{Ludovica Cicci}
\author[a]{Shuang Qian}
\author[a,b]{Steven A. Niederer}

\cortext[cor1]{Corresponding author}
\address[a]{National Heart and Lung Institute, Imperial College London, London, UK}
\address[b]{The Alan Turing Institute, London, UK}

\begin{abstract}
Cardiac electrophysiology (CEP) simulations are increasingly used for understanding cardiac arrhythmias and guiding clinical decisions. However, these simulations typically require high-performance computing resources with numerous CPU cores, which are often inaccessible to many research groups and clinicians. To address this, we present TorchCor, a high-performance Python library for CEP simulations using the finite element method on general-purpose GPUs. Built on PyTorch, TorchCor significantly accelerates CEP simulations, particularly for large 3D meshes. The accuracy of the solver is verified against manufactured analytical solutions and the $N$-version benchmark problem. TorchCor is freely available for both academic and commercial use without restrictions.
\end{abstract}


\end{frontmatter}

%\linenumbers
\section{Motivation and significance}
\label{sec:motivation}

Cardiac arrhythmias are a major and often life-threatening clinical challenge, representing a leading cause of cardiovascular morbidity and mortality \cite{srinivasan2018sudden, kingma2023overview, huikuri2001sudden, zipes2006neural}. They often result from pathological interactions between dysfunctions at the cellular, tissue, and organ scales. Understanding these mechanisms typically requires multi-scale models and simulations to isolate and study their underlying causes. Consequently, cardiac electrophysiology (CEP) simulations are increasingly important tools for understanding arrhythmia mechanisms and guiding effective clinical decisions \cite{gsell2024forcepss, plank2021opencarp, trayanova2024computational}. These critical simulations, which widely employ numerical methods such as the Finite Element Method (FEM) to model electrical signal propagation through cardiac tissue, typically require high-performance computing (HPC) resources with numerous CPU cores. This substantial computational requirement has historically created a significant barrier, making these sophisticated, high-fidelity simulations inaccessible to many research groups and clinicians who lack access to large computational clusters. Ultimately, this bottleneck severely limits the throughput of studies and hampers direct clinical integration. 

This challenge of running CEP simulations has driven the research community to develop various libraries and tools to facilitate and streamline the process. A key representative is openCARP \cite{plank2021opencarp}, which provides a unified simulation environment. Built upon PETSc \cite{petsc-web-page}, a popular numerical solver for partial differential equations (PDEs), openCARP is designed to run on cluster machines, utilising parallel CPU architectures to speed up the simulation.

With modern graphics processing units (GPUs) becoming increasingly affordable and accessible on personal computers, recent studies \cite{kaboudian2024fast,mena2015gpu,berg2025monoalg3d} have also explored methods to accelerate cardiac modelling leveraging this technology. The work in \cite{mena2015gpu} focused on a GPU-accelerated solver utilising FEM for spatial discretisation. Their software, developed entirely in CUDA, implemented both fully explicit and semi-implicit solvers for the monodomain model using operator splitting. Another notable development in HPC for CEP is MonoAlg3D \cite{berg2025monoalg3d}, an open-source, GPU-based solver designed for scalable simulations. Distinct from FEM-based tools, MonoAlg3D utilises the finite volume method to solve the monodomain equation. Featuring parallel dispatching via Message Passing Interface to effectively utilise GPU clusters for high-throughput simulation studies, it achieved a substantial speedup over a CPU solution.

Beyond these traditional numerical solvers, the computational biophysics community is rapidly embracing data-driven modelling, specifically techniques like Physics-Informed Neural Networks (PINNs) \cite{raissi2019physics, kovachki2023neural}  and neural operators \cite{lu2021learning, azizzadenesheli2024neural}. These AI-augmented methods, which integrate physical laws directly into neural network architectures, are almost exclusively developed within deep learning frameworks like PyTorch \cite{paszke2019pytorch}. However, incorporating these novel methods alongside high-fidelity FEM solvers has historically been hindered by architectural friction, as FEM codes and machine learning frameworks typically reside in separate, often incompatible, software ecosystems. This incompatibility makes direct comparison and effective integration complex, a critical roadblock for hybrid modelling approaches intended to deliver superior predictive accuracy and speed.

This challenge is compounded by a fundamental and irreversible shift in the broader HPC landscape. Scientific applications, including computationally intensive domains like fluid dynamics, are rapidly transitioning from purely CPU-based architectures to accelerated computing to leverage new, parallel hardware like GPUs \cite{sathyanarayana2025high, guerrero2025python}. This transition is driven by the fact that the limiting factor for modern HPC is no longer compute power but rather power budget. This is evidenced by the world’s three publicly listed exascale systems, namely, El Capitan \cite{elcapitan2024}, Frontier \cite{atchley2023frontier}, and Aurora \cite{allen2025aurora}, which are all CPU and GPU designs. The next generation of exascale machines will be GPU-based \cite{carmin2025assessment}, and thus developing more sustainable and efficient GPU solvers will be essential to fully exploit these emerging architectures. Moreover, the necessity for sustainable and energy-efficient computing is now a primary design constraint for next-generation machines. GPUs, by design, are fundamentally more energy efficient than CPUs, which is why GPU-accelerated systems consistently dominate the Green500 efficiency rankings, which explicitly measure performance per Watt (perf/W) \cite{lannelongue2023greener,feng2007green500}. This global need for energy-efficient computing is also a key driver for developing and deploying more sustainable and performant GPU-based solvers.

To bridge this gap, we introduce TorchCor, a high-performance Python library for GPU-accelerated CEP simulations using the FEM. TorchCor is built entirely upon PyTorch, which provides a highly optimised backend for efficient tensor operations on GPUs, including a specialised, production-grade toolset for handling sparse matrices, which are essential to FEM. This enables TorchCor to automatically benefit from ongoing low-level performance enhancements developed by the machine learning industry.

The core contributions of our software package are summarised as follows:

\begin{itemize}
    \item \textbf{GPU-Accelerated FEM}: TorchCor implements a GPU-accelerated FEM solver for the monodomain equation, providing a substantial increase in speed and strong-scaling performance compared to traditional parallel CPU solvers for large 3D meshes.
    \item \textbf{Rigorous Verification}: The accuracy and robustness of the solver are rigorously confirmed using a two-tiered verification strategy: comparing numerical output to manufactured analytical solutions to verify core mathematical correctness, and verifying the full non-linear system against the $N$-version benchmark problem \cite{niederer2011verification} alongside state-of-the-art FEM solvers.
    \item \textbf{Bridging Ecosystems for Hybrid Modelling}: By residing natively within the PyTorch ecosystem, TorchCor eliminates the historical architectural friction between traditional FEM solvers and data-driven methods. This unified platform establishes a critical testbed for methodological advancement, enabling the direct comparison and integration of high-fidelity mechanistic models with emerging AI-augmented techniques, such as PINNs or neural operators. 
    \item \textbf{Enhanced Accessibility and Usability}: Designed as a pure Python library with an intuitive API, TorchCor lowers the barrier to entry for researchers and clinicians, making high-throughput cardiac simulations feasible on local desktop workstations or easily integrated into clinical imaging environments like CemrgApp \cite{razeghi2020cemrgapp} and EP Workbench \cite{bodagh2025openep}.
\end{itemize}

This paper is organised as follows. Section~\ref{sec:description} describes the core functionalities, implementation details, and computational algorithms of TorchCor. Section~\ref{sec:illustrative} presents two tests used to verify the accuracy of the solver, followed by a discussion of its design and recommended usage for various purposes. We also evaluate the performance of our GPU solver against a state-of-the-art CPU-based simulator. Section~\ref{sec:impact} discusses the impact of our library on the machine learning community and its integration into other software for facilitating CEP simulations. Finally, Section~\ref{sec:conclusion} concludes the paper.

\section{Software description}
\label{sec:description}

TorchCor is a pure Python library that leverages the PyTorch framework as its computational backend. Its architecture is designed for ease of use, allowing users to define and run complex CEP simulations with a simple and intuitive API and providing comprehensive control over the simulation environment. The core components include a mesh loader, an FEM assembler for mass and stiffness matrices, a high-performance FEM linear solver, and an implementation for common ionic models, all of which are orchestrated to execute seamlessly on a GPU device.

\subsection{Software functionality}

The core functionality of TorchCor is realised through a logical progression of steps that define and execute the simulation based on the monodomain model \cite{geselowitz1983bidomain}. Users start by specifying the GPU device on which the simulation is executed and defining the temporal parameters, namely the total simulation time and the discrete time step. 

The user then specifies the ionic model describing the cell action potential. Several commonly used ionic models are available. Each ionic model is implemented as a Python class where the constants and state variables are represented by class attributes that users can modify in accordance with their simulation requirements.  

TorchCor implements functionalities that allow the computational domain, described as unstructured triangulations (triangular or tetrahedral elements), to be read from three input files: one for the 3D coordinates of the mesh vertices, another for the mesh connectivity, and finally one that specifies the fibre directions. The three files are expected to be located in the same folder and with the correct file extensions. 

%In TorchCor, spatial material properties can be heterogeneous, so users can assign specific conductivities to different heart regions. While the monodomain equation uses a conductivity tensor, the inputs allow for specifying values that reflect elementwise longitudinal and transverse conductivities of a tissue, for simulating anisotropic electrical conduction. 
In TorchCor, users can define varying electrical conductivities across different regions of the heart to reflect heterogeneous tissue properties. The conductivity tensor that characterises the monodomain equation is obtained by specifying the local (elementwise) values of the longitudinal and the transverse conductivities.

We allow users to apply any number of stimuli. Each stimulus is associated with a file containing the indices of the spatial coordinates of the stimulus region. The starting time, the stimulus duration, and its intensity are the required parameters to be specified for a stimulus to be successfully applied. 

The final step involves initialising the GPU-based FEM solver and executing it on the designated GPU. The linear system generated by FEM discretisation is solved by a Preconditioned Conjugate Gradient (PCG) method, with a Jacobi preconditioner.
The user has the flexibility to control details of the solver, such as the absolute and relative tolerances and the maximum number of iterations. The resulting action potential for all nodes can be saved at a fixed interval for downstream tasks or virtualisation purposes.     

During runtime, TorchCor has the option to record the local activation (LAT) and repolarisation  (LRT) times for each node. 
The LAT is defined as the moment when the transmembrane potential first becomes positive, whereas the LRT denotes the instant when the potential subsequently returns to its resting value below -70 with a negative time derivative, marking the completion of a full action potential cycle. 

\subsection{Software implementation}
The monodomain equation \cite{geselowitz1983bidomain}, which governs electrical signal propagation in cardiac tissue, is described by the following set of equations:
%The computational core of TorchCor is the GPU implementation of FEM with linear shape functions for the monodomain equation \cite{geselowitz1983bidomain}: 
%, which governs electrical signal propagation in cardiac tissue. The strong form of the equation is given by:

\begin{equation}
    \label{eq:monodomain_strong}
    \begin{aligned}
    &\chi \left( C_m \frac{\partial V}{\partial t} + I_{\text{ion}}(V, \mathbf{u}) \right) 
    = \nabla \cdot \left( \mathbf{\sigma} \nabla V \right) + \chi I_{\text{stim}} \quad & \text{in } \Omega \times (0, T]\\
    &\frac{\partial\mathbf{u}}{\partial t} = \mathbf{g}(V,\mathbf{u}) \quad & \text{in } \Omega \times (0, T]\\
    &\mathbf{n} \cdot (\mathbf{\sigma} \nabla V) =0 & \text{in } \partial\Omega \times (0, T]\\
    &V(\mathbf{x}, 0) = V_0(\mathbf{x}) & \text{in } \Omega\\
    &\mathbf{u}(\mathbf{x}, 0) = \mathbf{u}_0(\mathbf{x}) & \text{in } \Omega
    \end{aligned}
\end{equation}
%with the no-flux boundary condition $\mathbf{n} \cdot (\mathbf{\sigma} \nabla V) = 0$ on the boundary $\partial \Omega$, and an initial condition $V(\mathbf{x}, 0) = V_0(\mathbf{x})$. 
Here $V$ is the transmembrane potential, $\mathbf{\sigma}$ is the positive definite conductivity tensor, $I_{\text{stim}}$ is an external stimulus current, $\chi$ is the surface-area-to-volume ratio, $C_m$ is the membrane capacitance, $I_{\text{ion}}$ is the ionic current density determined by the chosen cell model and depends on the vector of state variables $\mathbf{u}$ that follow a temporal dynamics $\mathbf{g}$. A comprehensive list of these variables and their units is provided in Table~\ref{tab:variables}.

TorchCor incorporates well-established electrophysiological models, including the Ten Tusscher–Panfilov model for human ventricular cells \cite{ten2006alternans} and the Courtemanche–Ramirez–Nattel (CRN) model \cite{courtemanche1998ionic}, which provides a biophysically detailed representation of the human atrial action potential.

\begin{table}[H]
    \centering
    \caption{List of variables and their units}
    \label{tab:variables}
    \begin{tabular}{l l l}
        \toprule
        \textbf{Variable} & \textbf{Name} & \textbf{Unit} \\ 
        \midrule
        %$\Delta x$ & mesh discretization  & millimetre (mm) \\ 
        %$\Delta t$ & temporal unit  & millisecond (ms) \\ 
        $\Omega$ & space  & millimetre (mm) \\ 
        $t$ & time  & millisecond (ms) \\        
        $V$ & membrane potential  & millivolt (mV) \\ 
        $I_{ion}$ & transmembrane ionic current & $\mu A \, \text{mm}^{-2}$ \\ 
        $I_{stim}$ & stimulus intensity & $\mu A \, \text{mm}^{-2}$ \\ 
        $\chi$ & surface area to volume ratio  & $\text{mm}^{-1}$ \\ 
        $C_m$ & membrane capacitance & $\mu F \, \text{mm}^{-2}$ \\ 
        $\sigma$ & conductivities  & $S \, \text{m}^{-1}$ \\ 
        \bottomrule
    \end{tabular}
\end{table}

System (\ref{eq:monodomain_strong}) is discretised in space using linear Finite Elements and in time using the $\theta$~-~method \cite{heidenreich2010adaptive}:
\begin{equation}
\begin{aligned}
&\mathbf{u}^{k+1} = \mathbf{u}^{k} + \Delta t \mathbf{g}(\mathbf{V}^{k},\mathbf{u}^{k})   \\
    &\left(\chi C_m \mathbf{M}  + \theta\Delta t\mathbf{K} \right)\mathbf{V}^{k+1} =\left(\chi C_m \mathbf{M}  + (1-\theta)\Delta t\mathbf{K} \right)\mathbf{V}^{k}+\\
    &\quad\quad\quad\quad\quad\quad\quad\quad\quad 
    +\chi\Delta t \mathbf{M} \left(  \mathbf{I}_{\text{stim}} - \mathbf{I}_{\text{ion}} \left(\mathbf{V}^{k}, \mathbf{u}^{k+1}\right)   \right),
\end{aligned}  \label{eq:sysdiscrete}  
\end{equation}
where $\mathbf{V}^{k}, \mathbf{u}^{k}$ are the vectors of nodal potentials and state variables at time $k$, and $\mathbf{M}$ and $\mathbf{K}$ are the global mass and stiffness matrices, respectively.
Matrices $\mathbf{M}$ and $\mathbf{K}$ are of size $n \times n$ (where $n$ is the number of nodes) and are sparse. TorchCor leverages PyTorch's specialised sparse matrix formats, primarily the Coordinate (COO) format for sparse matrix construction and manipulation, and the Compressed Sparse Row (CSR) format for fast iterative matrix-vector multiplications on the GPU. To further enhance the efficiency of matrix-vector products, the matrix structure is optimised by reordering its elements using the Reverse Cuthill-McKee (RCM) algorithm \cite{cuthill1969reducing} to minimise the bandwidth.

% For a given cardiac geometry discretised into a mesh (supporting 2D, 3D surface, and 3D volume elements), we apply the FEM with linear basis functions to discretise the monodomain equation in space. This process involves assembling a mass matrix $\mathbf{M}$, which represents the capacitance and ionic current terms, and a stiffness matrix $\mathbf{K}$, which represents the diffusion term. Both matrices are of size $n \times n$, where $n$ is the number of nodes in the mesh. For any realistic cardiac mesh, these matrices are extremely sparse, meaning the vast majority of their entries are zero. To optimise memory and computational efficiency, TorchCor leverages PyTorch's specialised sparse matrix formats and operations. To be specific, we use COO to represent the sparse matrices and change them into CSR matrix for fast matrix and vector multiplication. To enhance the performance of iterative matrix-vector multiplications, the matrix structure is optimised by reordering its elements using the Reverse Cuthill-McKee algorithm \cite{cuthill1969reducing}.

%After spatial discretisation, we apply a finite difference approximation in time using the generalised-$\theta$ scheme. This transforms the partial differential equation into a sequence of linear systems, one for each time step $k+1$, of the form:

The algorithm first advances in time the state variables (first row of \eqref{eq:sysdiscrete}), constituting a system of ordinary differential equations using an explicit forward Euler scheme. Then, it solves the following linear system:
\begin{equation}
    \label{eq:cg}
    \mathbf{A} \mathbf{V}^{k+1} = \mathbf{b},
\end{equation}
%where $\mathbf{V}^{k+1}$ is the vector of unknown membrane potentials at the next time step. The system matrix $\mathbf{A}$ and the right-hand side vector $\mathbf{b}$ are defined as:
where:
\begin{align*}
    \mathbf{A} &= \chi C_m \mathbf{M} + \theta \Delta t \mathbf{K} \\
    \mathbf{b} &= \chi \mathbf{M}  \left(  C_m \mathbf{V}^k - \Delta t  I_{\text{ion}} + \Delta t I_{\text{stim}} \right) 
    - (1 - \theta) \Delta t \mathbf{K} \mathbf{V}^k.
\end{align*}
The temporal integration scheme is determined by the parameter $\theta$. As shown in Table~\ref{tab:schemes}, different $\theta$ values lead to different integration schemes. 
TorchCor defaults to the unconditionally stable Crank-Nicolson method ($\theta$=0.5), which offers second-order accuracy in time and provides a good balance between stability and accuracy. At each time step, the resulting linear system is solved using the PCG method, described in Algorithm~\ref{alg:pcg}. This is an iterative method well-suited for the large, sparse, symmetric positive-definite systems that arise in FEM. A Jacobi preconditioner is applied by default to accelerate convergence. 
To mitigate numerical instability from accumulated rounding errors during long simulations \cite{pikle2018gpgpu}, a common issue in iterative solvers using single precision, TorchCor performs all core computations in double precision (float64) by default.

\begin{table}[H]
    \centering
    \caption{Properties of time integration schemes}
    \label{tab:schemes}
    \begin{tabular}{c l l}
        \toprule
        $\theta$ & \textbf{Method} & \textbf{Stability} \\ 
        \midrule
        0 & Forward Euler & conditionally stable \\ 
        0.5 & Crank-Nicolson & unconditionally stable \\ 
        $\frac{2}{3}$ & Galerkin scheme & unconditionally stable \\ 
        1 & Backward Euler & unconditionally stable \\ 
        \bottomrule
    \end{tabular}
\end{table}

\begin{algorithm}[H]
\caption{Preconditioned Conjugate Gradient (PCG) for FEM}
\label{alg:pcg}
\begin{algorithmic}
\Require Symmetric positive-definite matrix $A$, right-hand side vector $b$, initial guess $x_0$, preconditioner $M$, absolute tolerance $\epsilon_a$, relative tolerance  $\epsilon_r$, maximum number of iterations $m$
\Ensure Approximate solution $x$

\State Compute $r_0 = b - A x_0$ 
\State Solve $M z_0 = r_0$ 
\State Set $p_0 = z_0$
\State Set $x = x_0$
\State Compute $\rho_0 = r_0^T z_0$
\State Compute $\|z_{k}\|$
\For{$k = 0, 1, 2, \dots$ $m-1$}
    \State Compute $q_k = A p_k$
    \State Compute $\alpha_k = \frac{\rho_k}{p_k^T q_k}$
    \State Update $x = x + \alpha_k p_k$
    \State Update $r_{k+1} = r_k - \alpha_k q_k$
    \If{$\|z_{k+1}\| < \epsilon_a$ or $\frac{\|z_{k+1}\|}{\|z_{k}\|} < \epsilon_r$}
        \State \textbf{break}
    \EndIf
    \State Solve $M z_{k+1} = r_{k+1}$
    \State Compute $\rho_{k+1} = r_{k+1}^T z_{k+1}$
    \State Compute $\beta_k = \frac{\rho_{k+1}}{\rho_k}$
    \State Update $p_{k+1} = z_{k+1} + \beta_k p_k$
\EndFor
\end{algorithmic}
\end{algorithm}

In PCG, Torchor uses the following second-order extrapolation to choose the initial guess for $\textbf{V}^{k+1}$:
%for the PCG solver at time step k is not arbitrary. Instead of using the previous solution $\textbf{V}^{k-1}$, TorchCor employs a linear extrapolation from the two previous time steps to provide a more accurate initial estimate
\[
\textbf{V}^{k+1} = 2\textbf{V}^{k} - \textbf{V}^{k-1}.
\]

%The ionic current, $I_{\text{ion}}$, is computed by solving a system of ordinary differential equations (ODEs) that constitute the cellular ionic model.
%TorchCor incorporates well-established electrophysiological models, including the Ten Tusscher–Panfilov model for human ventricular cells \cite{ten2006alternans} and the Courtemanche–Ramirez–Nattel (CRN) model \cite{courtemanche1998ionic}, which provides a biophysically detailed representation of the human atrial action potential. 

\section{Illustrative Examples}
\label{sec:illustrative}

\subsection{Verification of simulation solutions}
We tested the accuracy and robustness of our implementation through the following two standard verification procedures. The combination of these two methods, namely the manufactured solution and the $N$-Version benchmark \cite{niederer2011verification}, is critical because the former verifies the correctness of the core mathematical implementation, while the latter verifies that the full, non-linear reaction-diffusion system (including the ionic model) accurately reproduces results from established simulation platforms.

\subsubsection{Verification via Manufactured Solutions}
\label{sec:manufactured_verification}

We performed a verification of the TorchCor diffusion solver (i.e., imposing  $I_{ion}=0$ in Eq.~\eqref{eq:monodomain_strong} ) to confirm its core mathematical accuracy and convergence properties. This process compares the numerical output of the solver to a known, manufactured analytical solution, a standard technique that involves modifying the governing partial differential equation to force a predetermined smooth function ($w$) to be its exact solution.

We set the surface-area-to-volume ratio ($\chi$) and membrane capacitance ($C_m$) to 1, and the conductivity tensor ($\mathbf{\sigma}$) to an identity matrix. The simplified equation reads:

\begin{equation}
\frac{\partial V}{\partial t} - \nabla^2 V = 0.
\label{eq:monodomain_simplified}
\end{equation}
We defined the target manufactured solution as a decaying wave travelling across the domain:

\begin{equation}
w(x, y, t) = e^{-kt} \cos(\omega_1 x + \omega_2 y - \lambda t).
\label{eq:manufactured_solution}
\end{equation}
Since this function $w$ does not inherently satisfy the simplified homogeneous equation, we must introduce a corrective source term, $r(x,y,t)$, derived from the residual of the homogeneous ordinary differential equation (Eq.~\eqref{eq:monodomain_simplified}):
\begin{equation}
r(x,y,t) = \frac{\partial w}{\partial t} - \nabla^2 w.
\label{eq:residual}
\end{equation}
For the defined manufactured solution, the residual takes the form:

\begin{equation}
\begin{split}
r(x,y,t) &= e^{-kt} \left[ -k \cos(\omega_1 x + \omega_2 y - \lambda t) + \lambda\sin(\omega_1 x + \omega_2 y - \lambda t) \right] \\
&\quad + (\omega_1 ^2 + \omega_2 ^ 2) \cdot w(x,y,t).
\end{split}
\label{eq:residual_explicit}
\end{equation}
By incorporating this residual as a source term, we create a new constitutive equation that is guaranteed to have the desired manufactured solution:
\begin{equation}
\frac{\partial V}{\partial t} - \nabla^2 V = r(x,y,t).
\label{eq:constitutive_with_source}
\end{equation}
The TorchCor solver was executed using this new equation, and the difference between its numerical approximation and the analytical ground truth was measured. The simulation was performed on a square domain of $[0,1]\times[0,1]$ mm, discretised by a uniform $100\times 100$ grid and triangulated using Delaunay methods, resulting in $10,000$ nodes. Time integration utilised the Crank-Nicolson ($\theta=0.5$) scheme with a timestep of $\Delta t=0.01$ ms over a total duration of $T=50$ ms. For boundary conditions, we enforced Dirichlet conditions across the entire perimeter ($\partial\Omega$), where the numerical solution $V$ was fixed to the manufactured solution $w$ at all times ($V|_{\partial\Omega}=w|_{\partial\Omega}$). The initial condition was also set as $V(x,y,0)=w(x,y,0)$. The resulting linear system at each step was solved using the PCG method with a relative and absolute tolerance of $10^{-8}$. 

\begin{figure}[H]
\centering
\includegraphics[width=0.5\textwidth]{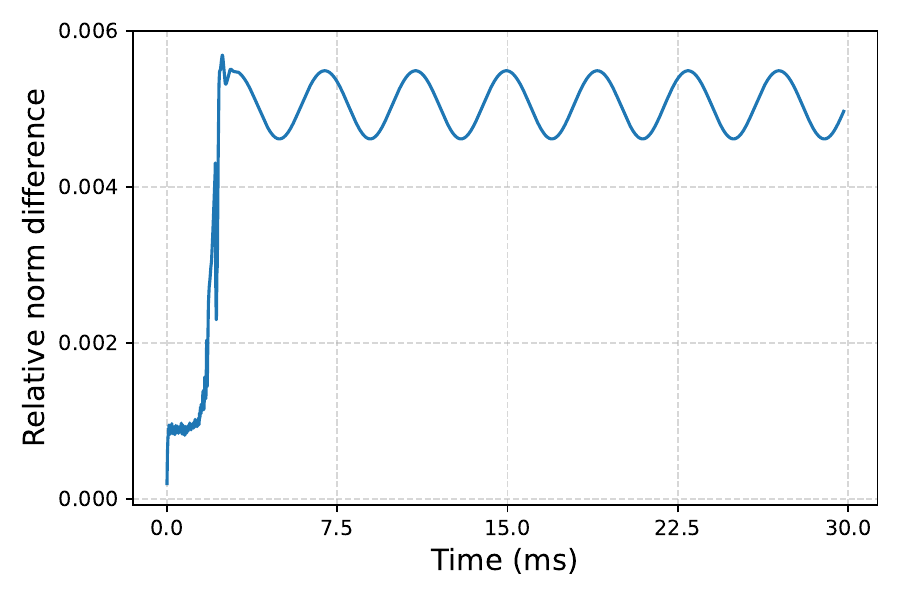}
\caption{The difference between the simulation solution from TorchCor and manufactured solutions over time.}
\label{fig:manufactured}
\end{figure}

As illustrated in Figure~\ref{fig:manufactured}, the error plot shows a typical initial transient phase where the error increases rapidly. This is due to the numerical solution adjusting from the perfect analytical initial condition to the steady-state accumulation of the numerical truncation error. Following this phase, the error stabilises into a low-magnitude, stable oscillation. This result confirms that the solver accurately tracks the periodic component of the decaying wave solution with both high precision and long-term stability.

\subsubsection{Verification via N-version benchmark}

We further verified TorchCor by implementing the $N$-version benchmark problem proposed by \cite{niederer2011verification}. This benchmark was introduced to address the significant challenge of code verification in cardiac tissue electrophysiology simulations. This verification ensures that a simulation code provides a reliable numerical solution of the underlying governing equations. It involves multiple simulation platforms solving an identical problem to establish a consensus gold-standard converged solution.

The benchmark is based on the monodomain equation with the Ten Tusscher-Panfilov ionic model for simulating electrical wave propagation in a simplified cuboidal cardiac tissue geometry. This geometry is a cube of size $20 \times 7 \times 3 \text{ mm}^3$, with fibres aligned along the $20 \text{ mm}$ (longitudinal) axis. The electrical wave is initiated by a constant current stimulus, $\mathbf{I_{\text{stim}}}$ applied to one square corner of this cube, where the corner point is referred to as $P_{1}$ (see Figure~1(a) in \cite{niederer2011verification}). Verification is primarily performed by evaluating the solution's convergence properties across varying spatial discretisations ($\Delta x \in \{0.1,\, 0.2,\, 0.5\}\ \text{mm}$) and a fixed temporal discretisation ($\Delta x$=0.005 ms). The standard metric for comparison is the activation time, defined as the time the membrane potential first passes through $0 \text{ mV}$ at the points on the diagonal line between two vertices $P_{1}$ and $P_{8}$ (see Figure~1(b) in \cite{niederer2011verification}). The simulation parameters are summarised in Table~\ref{tab:benchmark}.

\begin{table}[h]
    \centering
    \caption{List of variables and their units for the N-version benchmark}
    \label{tab:benchmark}
    \begin{tabular}{l l}
        \toprule
        \textbf{Parameter} & \textbf{Value} \\ 
        \midrule
        domain size & $20 \times 7 \times 3$ mm \\
        fibre orientation & aligned in the long axis, i.e., (1, 0, 0) \\
        $\Delta x$  & 0.5, 0.2, 0.1 mm \\ 
        $\Delta t$ & 0.005 ms \\ 
        $I_{ion}$ & Ten Tusscher-Panfilov \\
        $I_{stim}$ & 50 $\mu A \, \text{mm}^{-3}$ \\ 
        $\chi$ & 140 $\text{mm}^{-1}$ \\ 
        $C_m$ & 0.01 $\mu F \, \text{mm}^{-1}$\\ 
        $\sigma$ & longitudinal: 0.1334177 $S \, \text{m}^{-1}$\\ 
                 & transversal: 0.0173515 $S \, \text{m}^{-1}$\\
        \bottomrule
    \end{tabular}
\end{table}

\begin{figure}[H]
    \centering
    \includegraphics[width=0.5\textwidth]{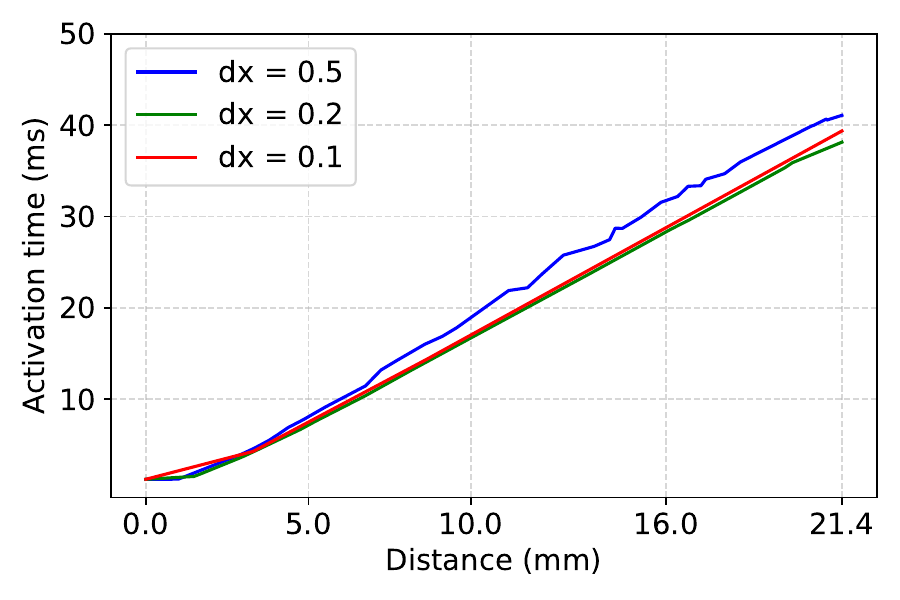}  % Adjust width to fit the text
    \caption{The performance of TorchCor on the $N$-version benchmark problem. It is comparable with Chaste and FEniCS solutions on this benchmark \cite{niederer2011verification}.}
    \label{fig:n_version}
\end{figure}

The simulation results from TorchCor are shown in  Figure~\ref{fig:n_version}.  The activation time from TorchCor is highly consistent with the stable reference codes -- such as Code A (Chaste \cite{pitt2009chaste}) and Code H (FEniCS \cite{mardal2007using}) in the original study \cite{niederer2011verification}) -- across the entire domain, demonstrating accurate conduction velocity along and across the fibre direction. Furthermore, an $L^2$-norm analysis (not shown here) of the difference in activation times between TorchCor's solution and the published gold-standard solution for the reference points $P_{1}$ and $P_{8}$ (i.e., the two diagonal vertices) confirms that TorchCor's performance is on par with the best-performing and most stable solvers evaluated in this benchmark, verifying that our GPU-based FEM implementation correctly solves the governing monodomain equations.

% \subsection{Best practices}

% Here, we demonstrate the designed use of our library for running CEP simulations on a biventricular mesh from \cite{biv_mesh}, with Tusscher \& Panfilov as the ionic model. 

% The interfaces and function names are designed in a self-explanatory and intuitive way. The \texttt{simulator} module provides the \texttt{Monodomain} model that contains the main functions for heavy computations on GPUs, while all implemented ionic models such as \texttt{TenTusscherPanfilov} are included inside the \texttt{ionic} module for modularity. 

% Considering that the spatial coordinates for the mesh are represented in units other than $mm$, the unit adopted in our library, we allow for unit conversion when loading the mesh. In this example, the coordinates are in $\mu m$ so we set \texttt{unit\_conversion} to 1000 to convert them to $mm$. 

% The regions that have the same conductivity are grouped to be assigned. Every stimulus is customisable with its own setting, including when it starts taking effect, how long it lasts and its intensity. 

% To start running the FEM solver, we suggest the default values for the PCG where tolerances are 0.00001 and the maximum number of iterations is 100. Setting \texttt{plot\_interval} to 10 and  \texttt{format} to vtk leads to the solutions being saved to Paraview file format for visualisation every 10 $mm$ till the end of the simulation. 

\subsection{Best Practices}

We demonstrate the implementation and designed usage of TorchCor by describing the setup for running a CEP simulation on a bi-ventricular mesh from \cite{biv_mesh}, using the Ten Tusscher-Panfilov ionic model.

The library's interfaces and function names are designed to be intuitive and self-explanatory, promoting ease of use and maintainability. The core functionality is divided into modular components. For instance, the \texttt{simulator} module provides the primary class for solving the chosen partial differential equation, i.e., the \texttt{Monodomain} model, which encapsulates the heavy computational routines optimised for GPUs. All implemented ionic models, including \texttt{TenTusscherPanfilov}, are organised within the \texttt{ionic} module. This modular separation simplifies the process of adding new ionic models to the library.

The simulation setup allows for flexible configuration. Recognising that mesh coordinates may be provided in various units, we incorporate an optional unit conversion factor during mesh loading. For instance, if the spatial coordinates of a mesh are expressed in microns instead of millimetres, the \texttt{unit\_conversion} parameter should be set to 1000 to correctly scale the mesh. Regions of the mesh that share the same conductivity can be grouped efficiently for unified parameter assignment. 
Furthermore, each stimulus applied to the tissue is highly customizable with individual settings for activation time, duration, and intensity.

To execute the FEM solver, we suggest setting the default values for the PCG solver. In typical use cases, the recommended tolerance is $10^{-5}$ and the maximum number of iterations is 100. For output visualisation, setting \texttt{plot\_interval} to 10 and \texttt{format} to \texttt{vtk} instructs the solver to save the solution every 10 ms in the VTK format \cite{vtkBook}, enabling post-simulation analysis.

\begin{lstlisting}[language=Python]
import torchcor as tc
from torchcor.simulator import Monodomain
from torchcor.ionic import TenTusscherPanfilov

tc.set_device("cuda:0")
simulation_time = 500
dt = 0.01

ionic_model = TenTusscherPanfilov(cell_type="ENDO", dt=dt)

mesh_dir = "/Data/ventricle/"
simulator = Monodomain(ionic_model, T=simulation_time, dt=dt)
simulator.load_mesh(path=mesh_dir, unit_conversion=1000)
simulator.add_conductivity([34, 35], 
                          il=0.5272, 
                          it=0.2076, 
                          el=1.0732, 
                          et=0.4227)
simulator.add_conductivity([44, 45, 46], 
                          il=0.9074, 
                          it=0.3332, 
                          el=0.9074, 
                          et=0.3332)
simulator.add_stimulus(f"{mesh_dir}/LV_sf.vtx", 
                       start=0.0, 
                       duration=1.0, 
                       intensity=100)
simulator.add_stimulus(f"{mesh_dir}/LV_pf.vtx", 
                       start=0.0, 
                       duration=1.0, 
                       intensity=100)
simulator.add_stimulus(f"{mesh_dir}/LV_af.vtx", 
                       start=0.0, 
                       duration=1.0, 
                       intensity=100)
simulator.add_stimulus(f"{mesh_dir}/RV_sf.vtx", 
                       start=5.0, 
                       duration=1.0, 
                       intensity=100)
simulator.add_stimulus(f"{mesh_dir}/RV_mod.vtx", 
                       start=5.0, 
                       duration=1.0, 
                       intensity=100)
simulator.solve(a_tol=1e-5, 
                r_tol=1e-5, 
                max_iter=100, 
                plot_interval=10, 
                verbose=True, 
                format="vtk")

\end{lstlisting}

\begin{figure}[h!]
    \centering
    \begin{minipage}{0.24\textwidth}
        \centering
        t=20 ms \\ % Time label for t=1
        \includegraphics[width=\textwidth]{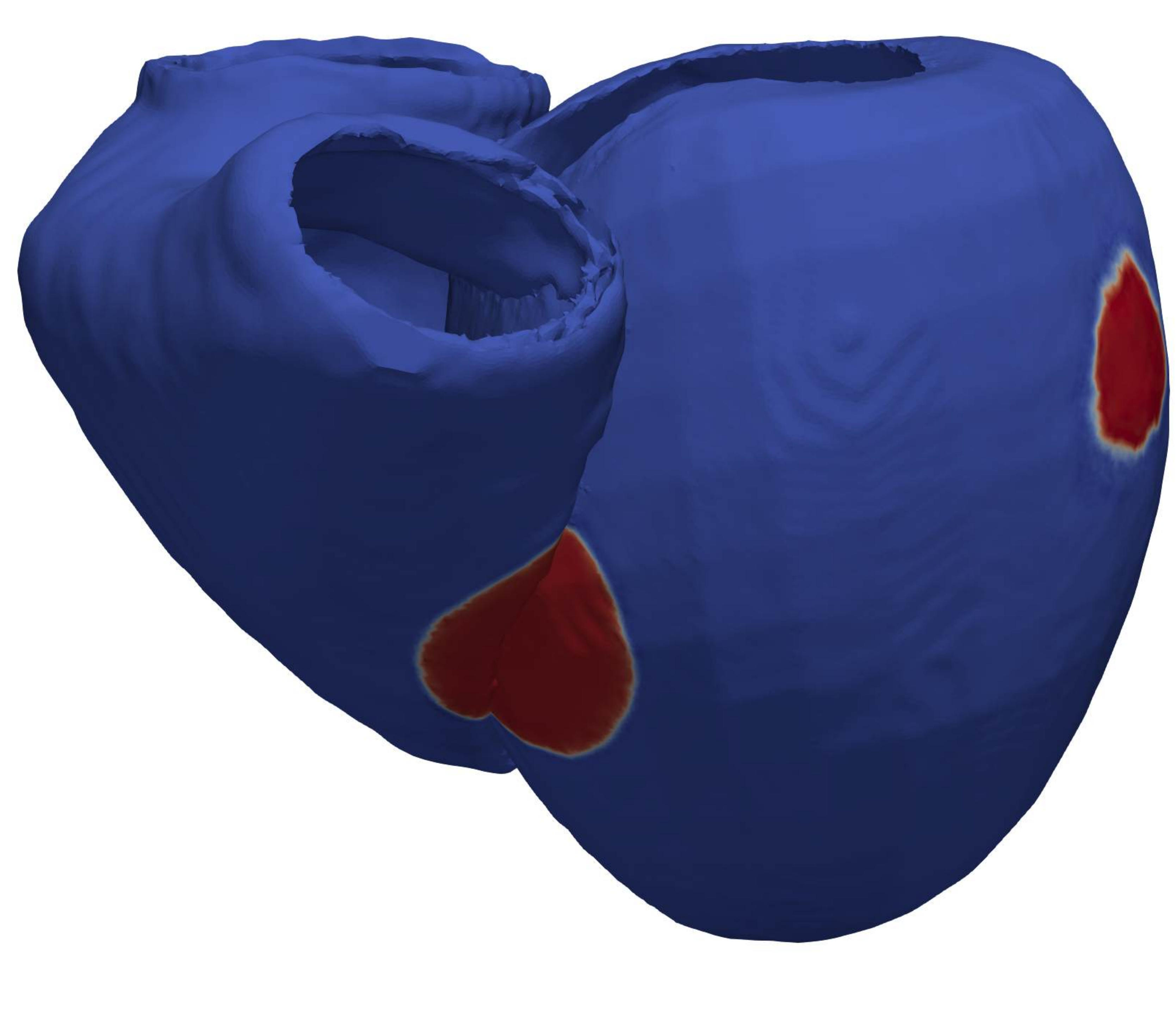} % Path to the first image
    \end{minipage}%
    \begin{minipage}{0.24\textwidth}
        \centering
        t=160 ms\\ % Time label for t=2
        \includegraphics[width=\textwidth]{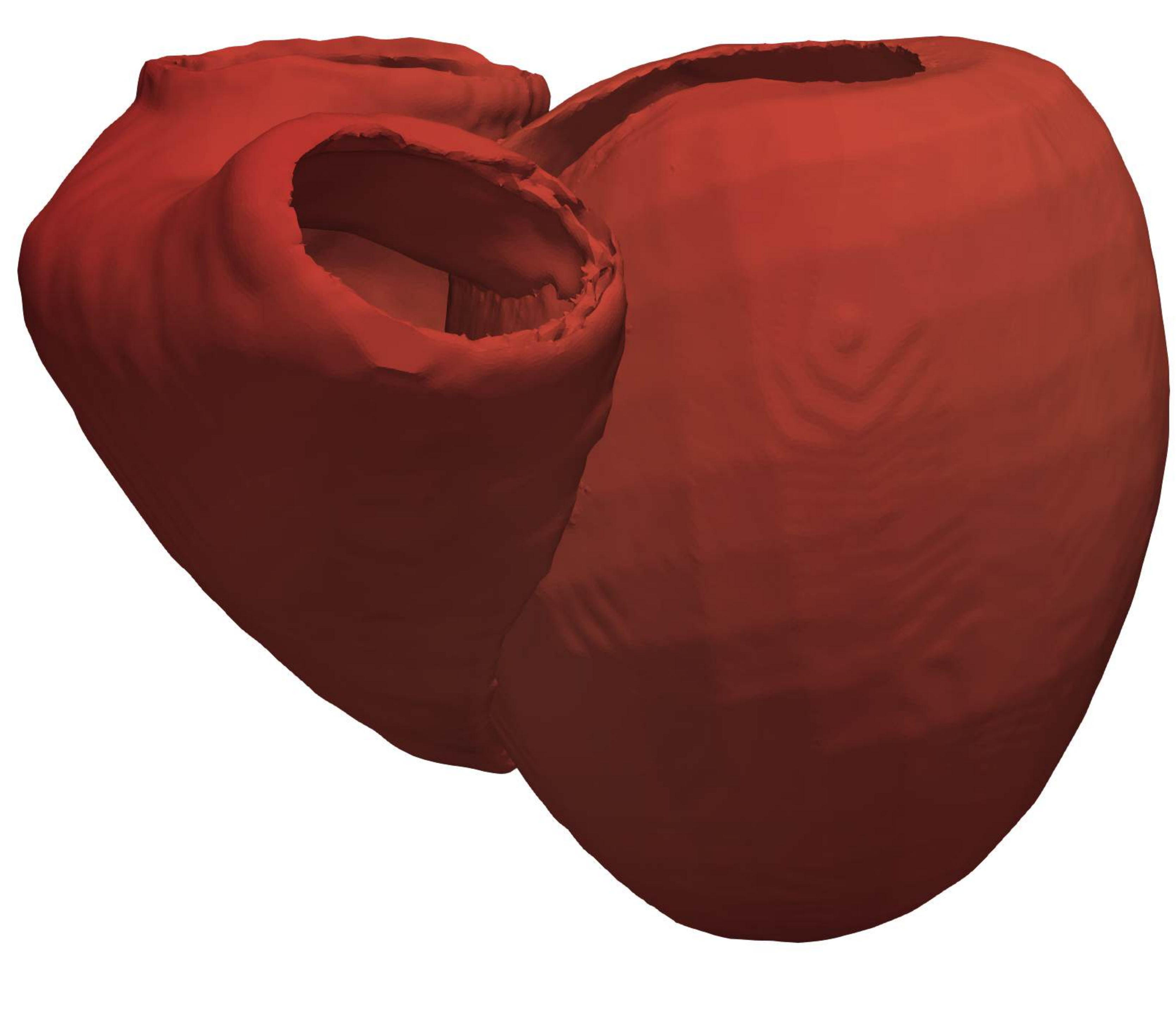} % Path to the second image
    \end{minipage}%
    \begin{minipage}{0.24\textwidth}
        \centering
        t=280 ms \\ % Time label for t=3
        \includegraphics[width=\textwidth]{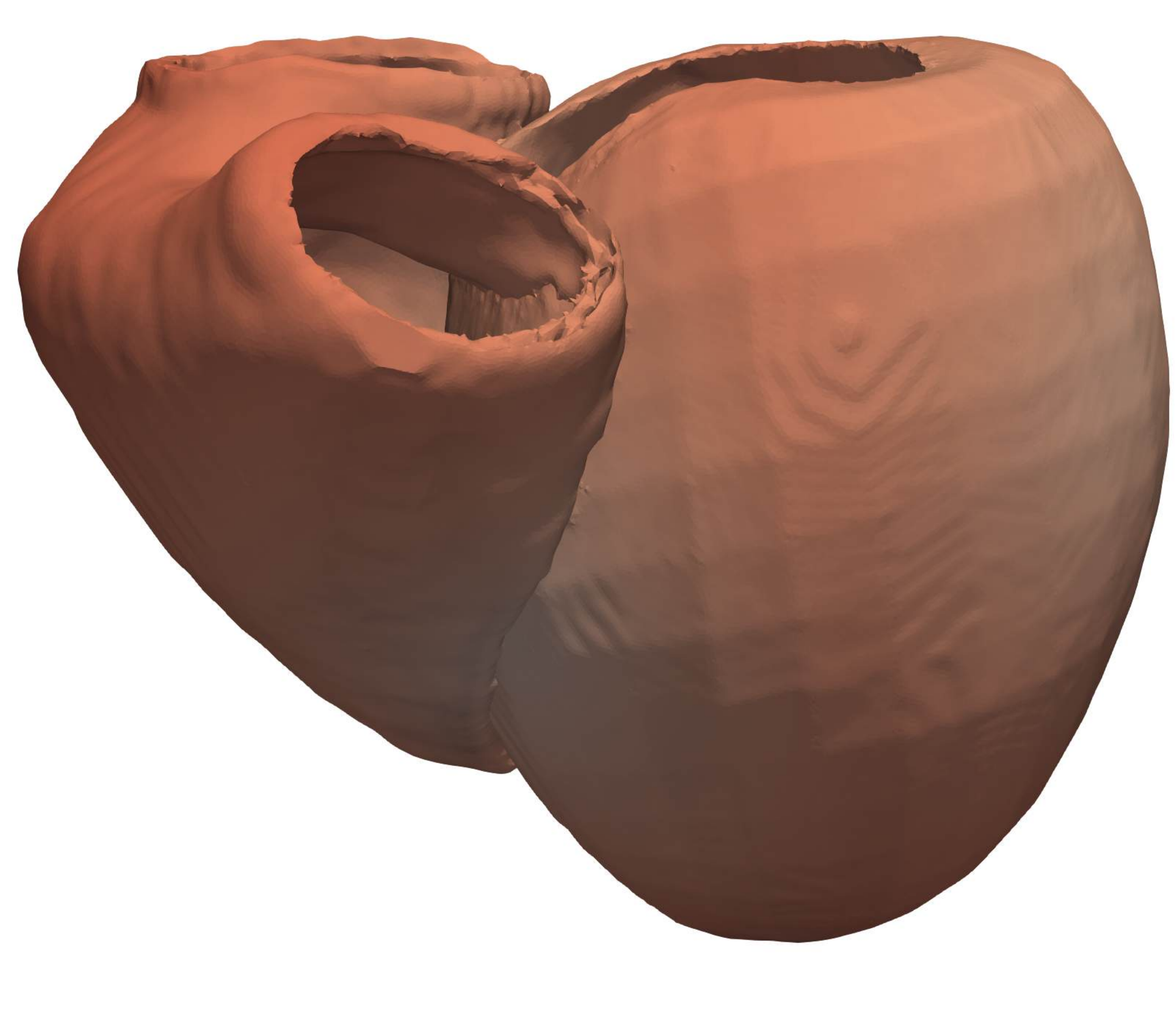} % Path to the third image
    \end{minipage}%
    \begin{minipage}{0.24\textwidth}
        \centering
        t=340 ms \\ % Time label for t=4
        \includegraphics[width=\textwidth]{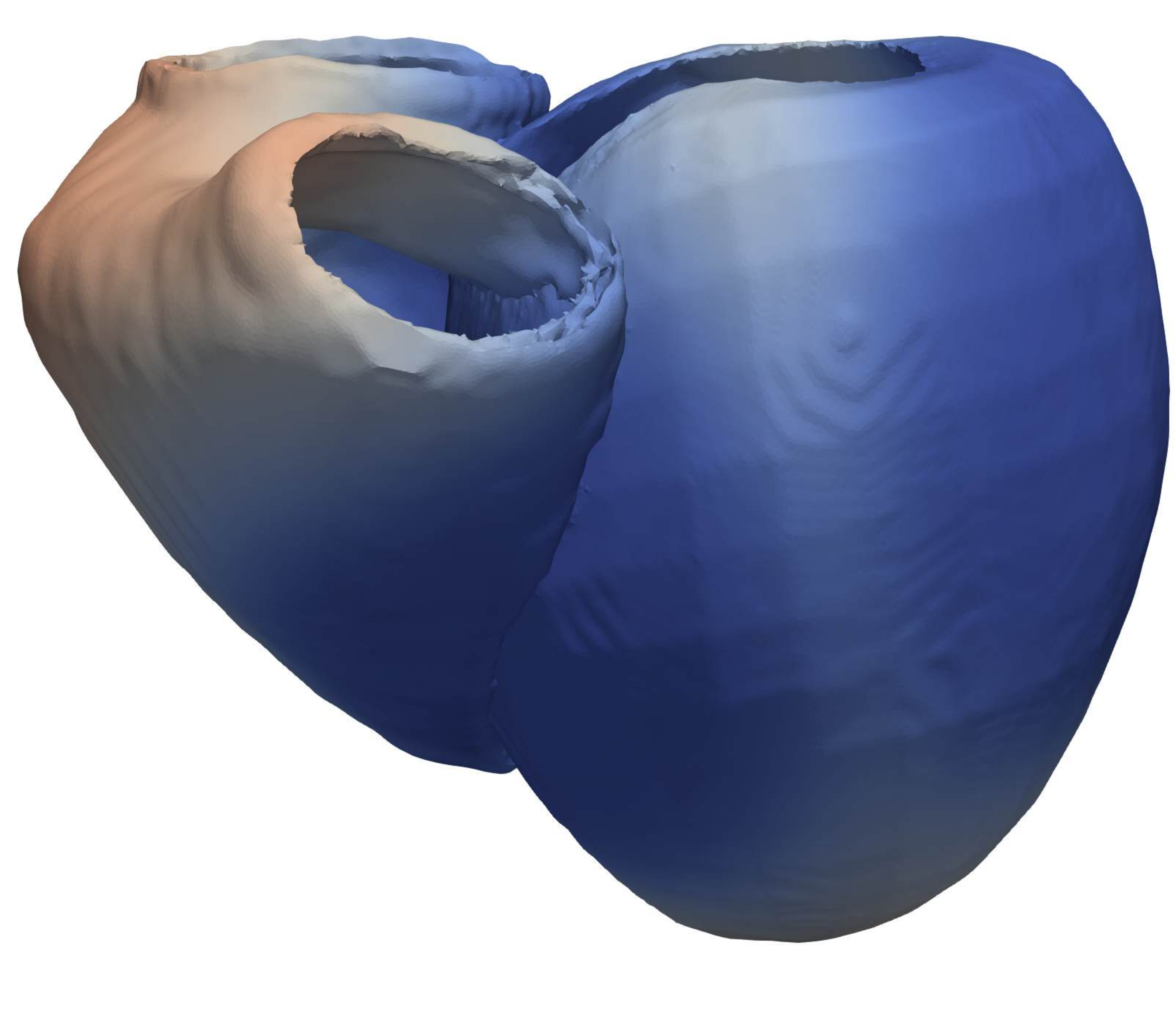} % Path to the fourth image
    \end{minipage}%
    \caption{Visualisation of simulation solutions on the bi-ventricle mesh at specified time.}
    \label{fig:biv}
\end{figure}

Figure~\ref{fig:biv} depicts the simulation results on the bi-ventricle mesh at four specified time points as the simulation progresses, illustrating the electrical signal propagation from the five initial stimuli. The simulation shows the entire organ undergoing the depolarisation and repolarisation phases within the 500 ms simulation time. For the complete solution data, please refer to our GitHub repository\footnote{TorchCor GitHub Repository: \url{https://github.com/sagebei/torchcor.git}}.

In the next example, we apply TorchCor to address a common challenge in cardiac research: managing high-throughput simulations. To demonstrate this capability, we utilise a cohort of one hundred patient-specific left atrium meshes sourced from the study by Roney et al. \cite{atrium_meshes}. The accompanying code snippet illustrates a Python-based workflow for efficiently running these large-scale cardiac electrophysiology simulations. Notably, we showcase the flexibility of the API by explicitly demonstrating that state variables and parameters of the ionic model can be programmatically reset to any justifiable user-specified value (e.g., for parameter sweeps or sensitivity analysis). While this script executes cases sequentially for clarity, it is inherently designed for deployment on HPC platforms. We suggest utilising array jobs on HPC clusters to parallelise this process, enabling each of the one hundred cases to be processed independently and efficiently on a dedicated GPU device.

\begin{lstlisting}[language=Python]
import torchcor as tc
from torchcor.simulator import Monodomain
from torchcor.ionic import ModifiedMS2v

tc.set_device("cuda:0")
simulation_time = 500
dt = 0.01

ionic_model = ModifiedMS2v(dt)
ionic_model.u_gate = 0.1
ionic_model.u_crit = 0.1
ionic_model.tau_in = 0.15
ionic_model.tau_out = 1.5
ionic_model.tau_open = 105.0
ionic_model.tau_close = 185.0

for case_id in range(100):
    case_name = f"Case_{case_id}"
    mesh_dir = f"/Data/atrium/{case_name}/"
    simulator = Monodomain(ionic_model, T=simulation_time, dt=dt)
    simulator.load_mesh(path=mesh_dir, unit_conversion=1000)
    simulator.add_conductivity(region_ids=[1, 2, 3, 4, 5, 6], 
                              il=0.4, 
                              it=0.4)
    simulator.add_stimulus(f"{mesh_dir}/{case_name}.vtx", 
                           start=0.0, 
                           duration=2.0, 
                           intensity=50)
    simulator.solve(a_tol=1e-5, 
                    r_tol=1e-5, 
                    max_iter=100, 
                    plot_interval=10, 
                    verbose=True,
                    format="vtk")
\end{lstlisting}

\begin{figure}[h!]
    \centering
    \begin{minipage}{0.24\textwidth}
        \centering
        t=10 ms \\ % Time label for t=1
        \includegraphics[width=\textwidth]{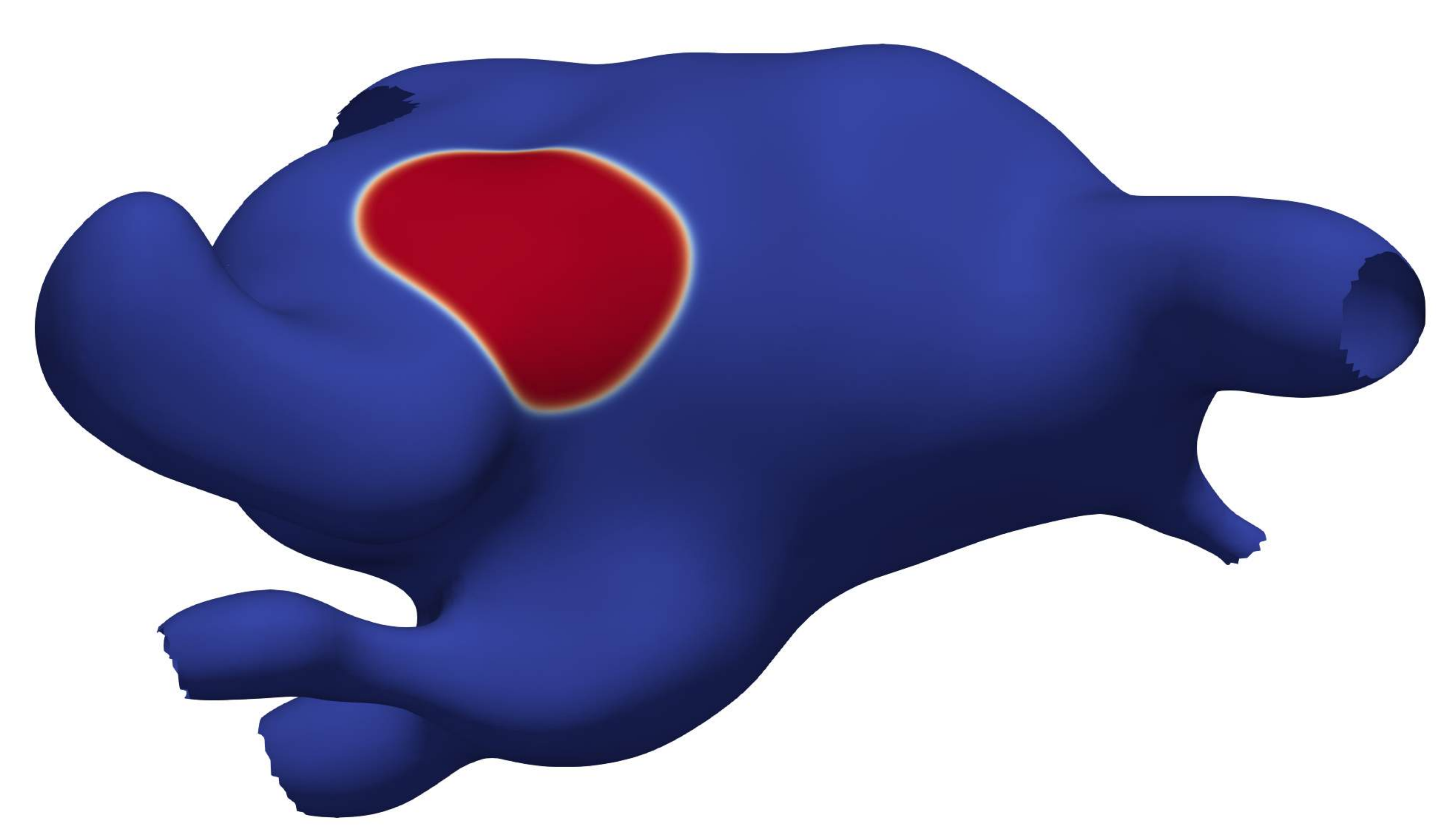} % Path to the first image
    \end{minipage}%
    \begin{minipage}{0.24\textwidth}
        \centering
        t=70 ms\\ % Time label for t=2
        \includegraphics[width=\textwidth]{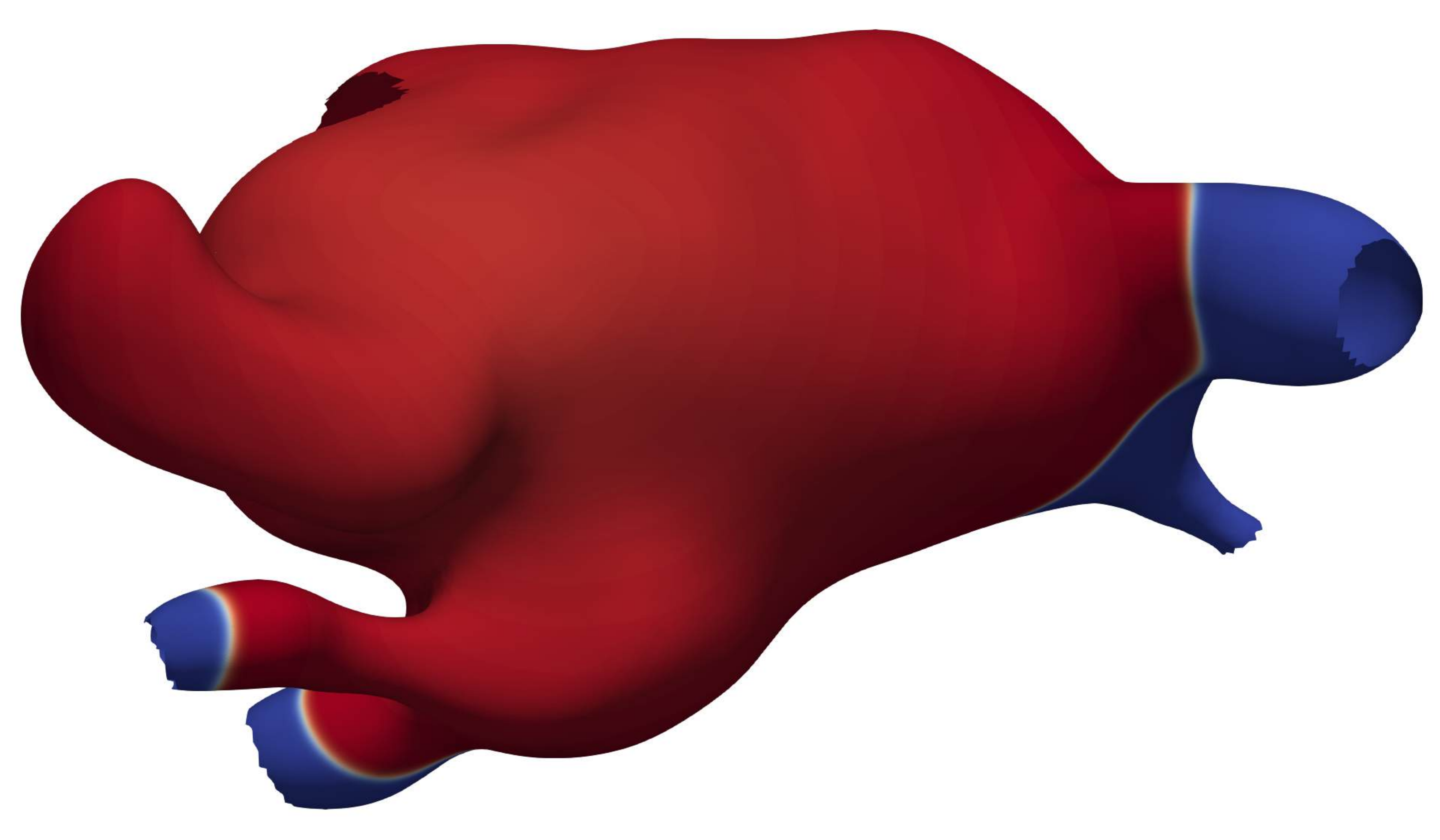} % Path to the second image
    \end{minipage}%
    \begin{minipage}{0.24\textwidth}
        \centering
        t=190 ms \\ % Time label for t=3
        \includegraphics[width=\textwidth]{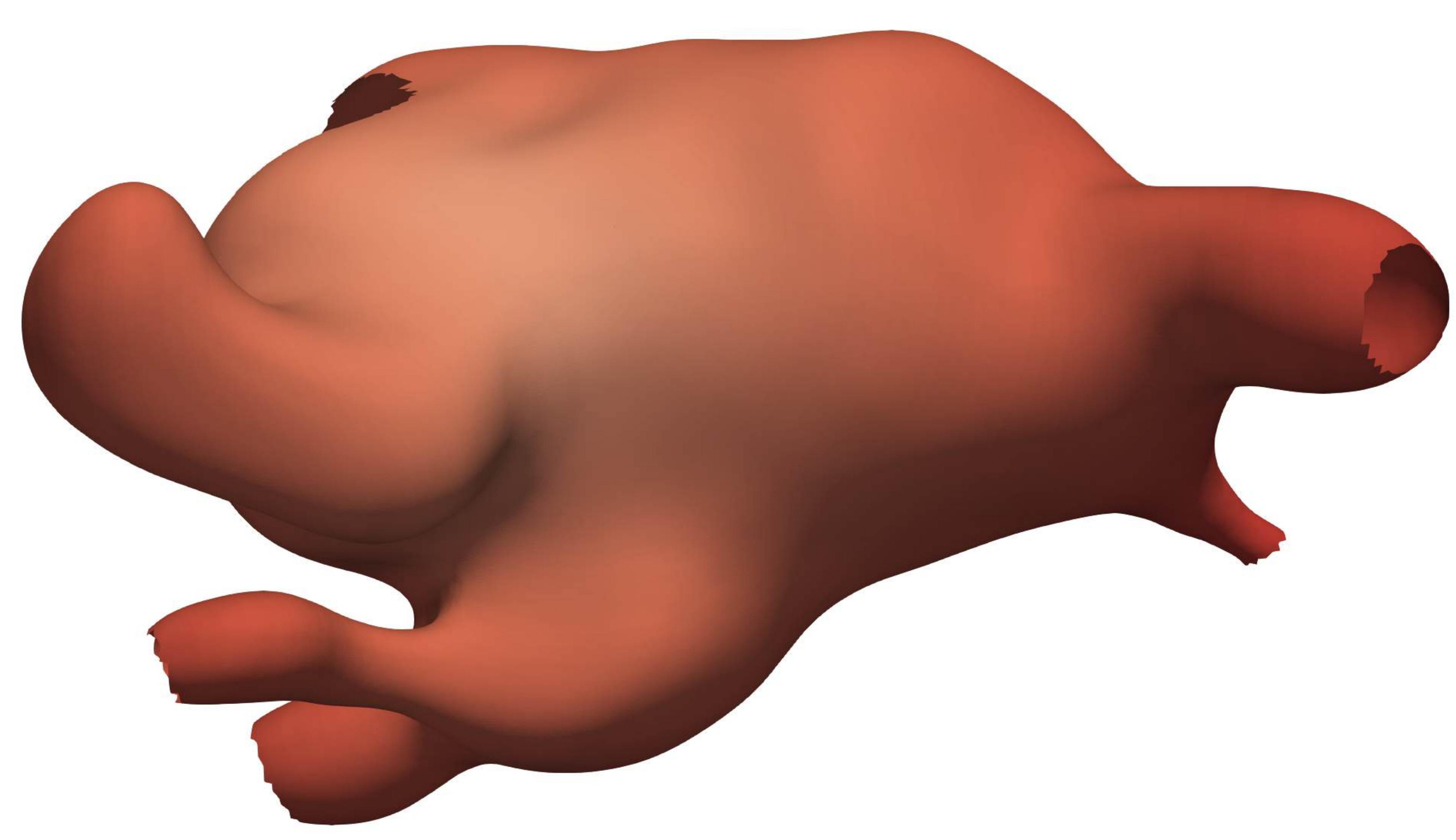} % Path to the third image
    \end{minipage}%
    \begin{minipage}{0.24\textwidth}
        \centering
        t=260 ms \\ % Time label for t=4
        \includegraphics[width=\textwidth]{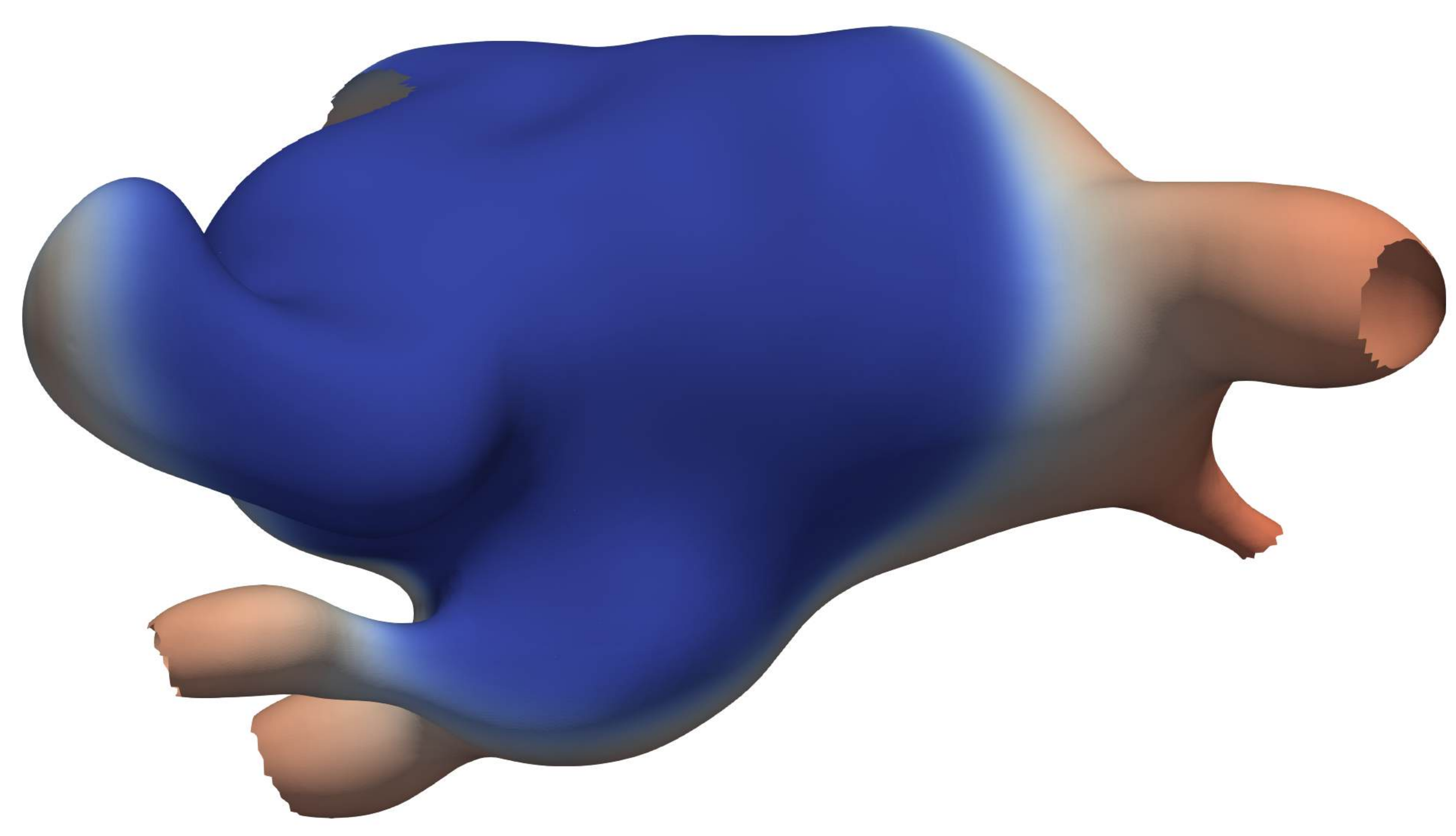} % Path to the fourth image
    \end{minipage}%
    \caption{Visualisation of the simulation solutions on a left atrium mesh with 660557 nodes at the specified times. }
    \label{fig:atrium}
\end{figure}

Figure~\ref{fig:atrium} displays the simulation results of one of these left atrium meshes with $660,557$ nodes. 

\subsection{Performance measurements}

\begin{figure}[H]
    \centering

    % --- First Row: 3D Volume Meshes ---
    \begin{subfigure}[t]{0.48\textwidth}
        \centering
        \includegraphics[width=\textwidth]{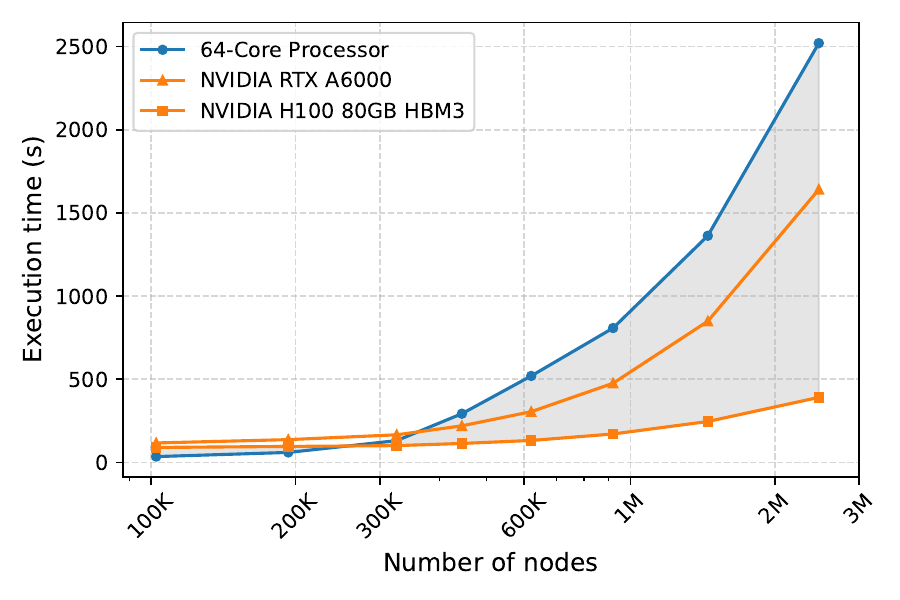}
        \caption{The execution time on cubic 3D volume meshes with varying numbers of nodes. The experiments on TorchCor were conducted only on NVIDIA RTX A6000 and H100 80GB HBM3 to showcase the full performance envelope of the GPU implementation on idealised geometry. The H100 represents the best-case performance (latest data centre GPU), while the RTX A6000 represents a strong professional workstation GPU for comparison.}
        \label{fig:volume_cpu_gpu}
    \end{subfigure}
    \hfill
    \begin{subfigure}[t]{0.48\textwidth}
        \centering
        \includegraphics[width=\textwidth]{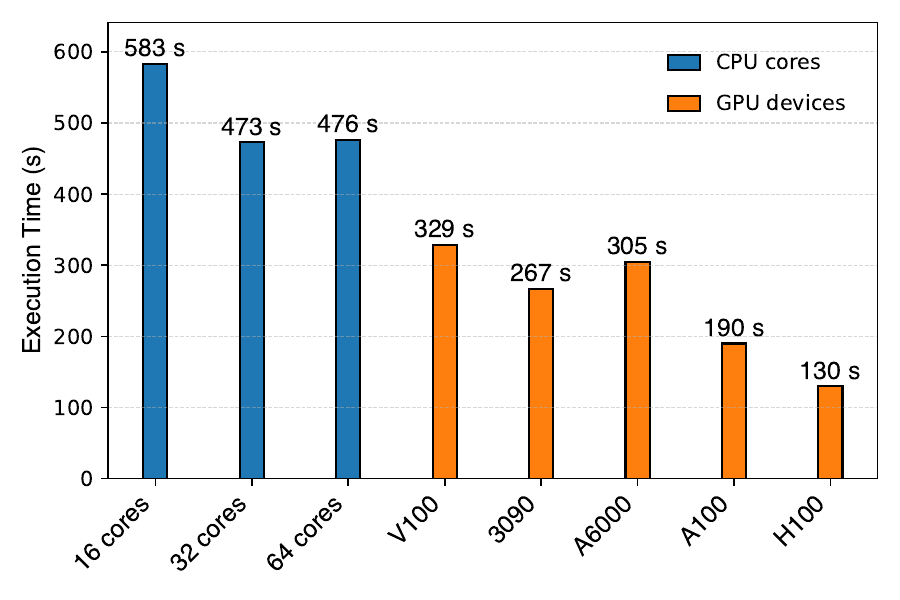}
        \caption{The execution time on the bi-ventricle 3D volume mesh with $637,480$ nodes across varying CPU core counts and different GPU devices. The GPUs tested here are NVIDIA Tesla V100, GeForce RTX 3090, RTX A6000, A100 80GB
PCIe, and H100 80GB HBM3}
        \label{fig:ventricle_cpu_gpu}
    \end{subfigure}

    \vspace{0.3em} % Add vertical space between the two rows

    % --- Second Row: 3D Surface Meshes ---
    \begin{subfigure}[t]{0.48\textwidth}
        \centering
        \includegraphics[width=\textwidth]{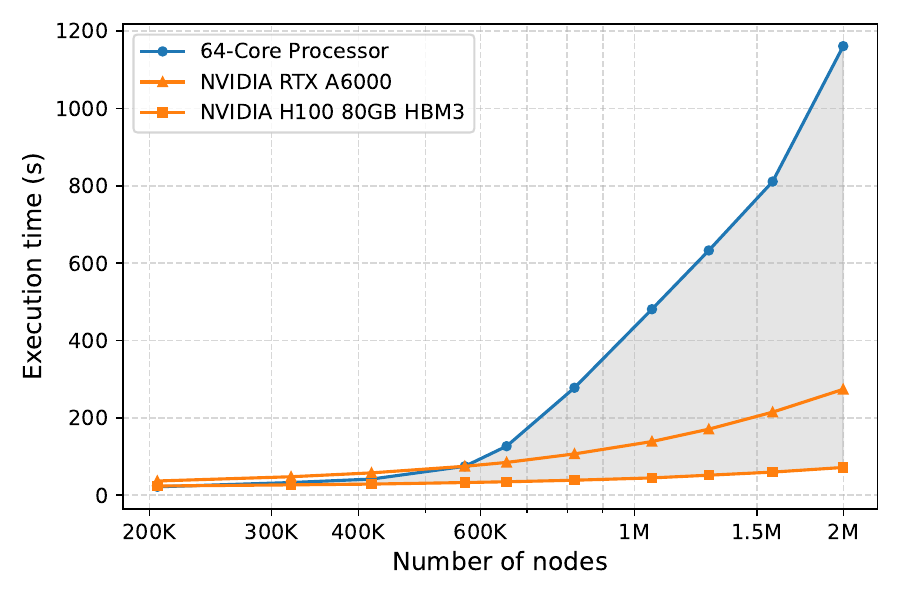}
        \caption{The execution time on a cubic 3D surface meshes with varying numbers of nodes.}
        \label{fig:surface_mms_cpu_gpu}
    \end{subfigure}
    \hfill
    \begin{subfigure}[t]{0.48\textwidth}
        \centering
        \includegraphics[width=\textwidth]{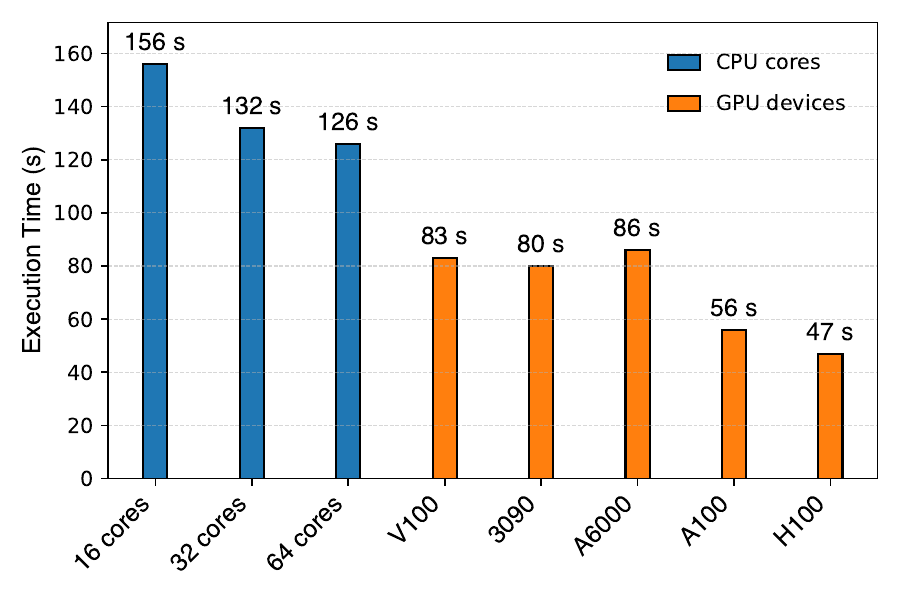}
        \caption{The execution time on the left atrium 3D surface mesh with $660,557$ nodes across varying CPU core counts and different GPU devices.}
        \label{fig:mms_cpu_gpu}
    \end{subfigure}

    \caption{Execution time for electrophysiology simulations. The top row (a, b) shows results on 3D volume meshes with the TenTussherPanfilov ionic model, and the bottom row (c, d) shows results on 3D surface meshes with a modified version of Mitchell-Schaeffer as the ionic model. Simulations use \texttt{openCARP} on CPU cores and \texttt{TorchCor} on GPU devices. The CPU used is an AMD Ryzen Threadripper 3990X 64-Core Processor.}
    \label{fig:performance_summary}
\end{figure}

We evaluate the performance of TorchCor on a wide range of modern GPUs (namely NVIDIA Tesla V100, GeForce RTX 3090, RTX A6000, A100 80GB PCIe,
and H100 80GB HBM3) using both idealised cube meshes with drastically increasing number of nodes and realistic patient-specific heart geometries for both surface and volume meshes. We compare these results against the parallelised CPU solver, openCARP, running on a 64-core desktop PC (AMD Ryzen Threadripper 3990X). All performance results are summarised in Fig. \ref{fig:performance_summary}.

We first assessed the scalability of TorchCor on idealised cube meshes (Fig. \ref{fig:volume_cpu_gpu} and \ref{fig:surface_mms_cpu_gpu}). On the 3D volume meshes (Fig. \ref{fig:volume_cpu_gpu}), for the smallest mesh size (around 100K nodes), the CPU-based solver had a slight initial advantage. However, the performance gap became substantial as the mesh size increased to 2.5M nodes. In this case, the 64-Core CPU exhibited the steepest increase in execution time, taking over 2500 s, while the NVIDIA RTX A6000 completed the simulation in approximately 1650 s. Crucially, the latest generation NVIDIA H100 80GB HBM3 proved to be the fastest, completing the simulation in only about 400 s. We conducted similar experiments on a 3D surface mesh with up to 2M nodes (Fig. \ref{fig:surface_mms_cpu_gpu}). The performance gap between the GPU and CPU solvers is even more pronounced for surface meshes, which typically have a higher ratio of surface area to volume, favouring the massive parallelisation capabilities of the GPU.

We also measured TorchCor performances on real patient-specific heart meshes used in recent research studies (Fig. \ref{fig:ventricle_cpu_gpu} and \ref{fig:mms_cpu_gpu}). For the bi-ventricle 3D volume mesh (around 637K nodes) (Fig. \ref{fig:ventricle_cpu_gpu}), we examined the effect of increasing the number of CPU cores. The simulation time is substantially reduced when more CPU cores are engaged in the computation up to 32 cores. However, we observed that the simulation with all 64 cores took a larger time than that with 32 cores. This is a common parallel computing bottleneck where the increasing inter-core communication cost surpasses the computational gains, leading to diminishing or negative returns on performance. In stark contrast, all tested GPUs delivered less than 330 s runtimes, with the H100 achieving a runtime of 130 s. A similar trend was observed for the left atrium 3D surface mesh (around 660K nodes) (Fig. \ref{fig:mms_cpu_gpu}), where TorchCor on the NVIDIA H100 reduced the execution time to approximately 47 s. These results strongly demonstrate the utility of TorchCor for rapidly simulating complex, realistic heart models across various modern GPU platforms.

In this work, we compared TorchCor performances against local high-performance workstations, which will inevitably continue to improve over time. The evolving balance between CPU performance and memory capacity will also influence future results. The comparisons presented here are therefore illustrative rather than definitive, as it is inherently difficult to make fair comparisons between CPU codes optimised for thousands of cores and GPU implementations.

\section{Impact}
\label{sec:impact}

PINNs have attracted significant attention for solving partial differential equations due to their ability to incorporate physical laws and constraints directly into the learning process. Their performance is often compared to that of traditional numerical methods, such as FEM \cite{grossmann2024can}. PINNs are typically implemented using deep learning frameworks like PyTorch and are optimised for training on GPU devices. Our library provides a platform to benchmark PINNs' performance, ensuring fair comparison by measuring execution time on identical GPU hardware.

TorchCor can be easily integrated into GUI applications such as CemrgApp \cite{razeghi2020cemrgapp}, an integrated environment for processing cardiovascular image data, whose overall design paradigm centres around offering interactive interfaces for external programs that run on Docker. To this end, we encapsulate TorchCor into a Docker image that runs on the command line in which all the simulation parameters are passed. It takes as input the mesh files and stores the solutions and solver statistic information in a designated directory. 

EP Workbench \cite{williams2021openep}, the graphical interface to OpenEP \cite{plank2021opencarp}, now includes TorchCor as the default GPU-based EP simulator. It allows users to configure simulation parameters via the GUI, which are then aggregated and saved to a configuration file that is sent to the local or remote GPU server.

\section{Conclusions}
\label{sec:conclusion}
We introduce TorchCor, a fast GPU-based finite element solver with a wide range of functionalities for large-scale electrophysiology simulations. By combining the numerical rigour of the finite element method with the computational efficiency of modern GPUs through PyTorch, it bridges the long-standing gap between accuracy, speed, and usability in cardiac modelling. The software delivers validated performance on realistic 3D geometries, scaling seamlessly from research desktops to data-centre GPUs, and its open Python architecture encourages transparent experimentation and reproducibility.

Beyond accelerating conventional simulations, TorchCor opens an avenue for unifying data-driven and physics-based modelling within a single framework. Its integration with deep-learning ecosystems enables direct comparison with neural operators and physics-informed neural networks, promoting methodological innovation across computational physiology and AI-augmented biophysics.

By making advanced electrophysiology simulations both fast and approachable, TorchCor lowers the barrier to entry for researchers and clinicians alike, fostering a more inclusive and agile community in cardiac modelling. Future developments will focus on extending support for bidomain formulations, adaptive time stepping, and multi-GPU parallelism, paving the way toward real-time, whole-heart simulations and next-generation digital twins of the human heart.

\section*{Declaration of competing interest}
The authors declare that they have no known competing financial interests or personal relationships that could have appeared to
influence the work reported in this paper.

\section*{Acknowledgments}
This work is supported by the NIHR Imperial Biomedical Research Centre (BRC) and by the British Heart Foundation Centre of Research Excellence (RE/24/130023)

\bibliographystyle{elsarticle-num} 
\bibliography{references}

% \section*{Required Metadata}
\section*{Current code version}

\begin{table}[H]
\begin{tabular}{|l|p{6.5cm}|p{6.5cm}|}
\hline
\textbf{Nr.} & \textbf{Code metadata description} & \textbf{Please fill in this column} \\
\hline
C1 & Current code version & v1.2 \\
\hline
C2 & Permanent link to code/repository used for this code version & \url{https://github.com/sagebei/torchcor.git} \\
\hline
C3  & Permanent link to Reproducible Capsule & \url{https://pypi.org/project/torchcor/} \\
\hline
C4 & Legal Code License   &  CC BY-NC 4.0 \\
\hline
C5 & Code versioning system used & git and GitHub \\
\hline
C6 & Software code languages, tools, and services used & Python 3.11.10, Docker \\
\hline
C7 & Compilation requirements, operating environments \& dependencies & torch 2.5.1, matplotlib 3.8.4
, scipy 1.14.1, numpy 1.26.4, pyvista 0.44.1\\
\hline
C8 & If available Link to developer documentation/manual &  \\
\hline
C9 & Support email for questions & bei.zhou@imperial.ac.uk \\
\hline
\end{tabular}
\caption{Code metadata (mandatory)}
\label{tah:version} 
\end{table}

\end{document}